\documentclass[graybox]{svmult}

\usepackage[numbers,square,sort&compress]{natbib}
\usepackage{varioref}

\usepackage{mathptmx}       
\usepackage{helvet}         
\usepackage{courier}        
\usepackage{makeidx}         
\usepackage{graphicx}        
\usepackage{multicol}        
\usepackage[bottom]{footmisc}

\usepackage{url}

\def\mb{M$_B$~}
\def\mbx{M$_B$}

\def\ha{H$\alpha$~}

\def\xha{H$\alpha$}

\def\sm{$\sl M_\odot$~}
\def\sma{$\sl M_\odot$}
\def\smayr{$\sl M_\odot$yr$^{-1}$}
\def\lb{L$_B$~}
\def\lbx{L$_B$}

\def\mahix{${\sl M}_{\rm H I}$}

\def\mahi{{${\sl M}_{\rm HI}$}~}

\def\mahilb{\mahix/\lb}
\def\mahilbx{\mahix/\lbx}

\def\xkms{km s$^{-1}$}
\def\hhx{H$_2$}
\def\hh{H$_2$~}
\def\hii{H II~}
\def\acox{$\alpha_{CO}$}
\def\aco{$\alpha_{CO}$~}

\begin{document}

\title*{Star forming dwarf galaxies.}
\titlerunning{Star forming dwarf galaxies}
\author{Nils Bergvall}
\institute{Department of Physics and Astronomy\\
Box 516\\
75120 Uppsala\\
nils.bergvall@astro.uu.se}
%
%
\maketitle

\abstract{abstract}
Star forming dwarf galaxies (SFDGs) have a high gas content and low metallicities, reminiscent of the basic entities in hierarchical galaxy formation scenarios. In the young universe they probably also played a major role in the cosmic reionization. Their abundant presence in the local volume and their youthful character make them ideal objects for detailed studies of the initial stellar mass function (IMF), fundamental star formation processes and its feedback to the interstellar medium. Occasionally we witness SFDGs involved in extreme starbursts, giving rise to strongly elevated production of super star clusters and global superwinds, mechanisms yet to be explored in more detail. SFDGs is the initial state of all dwarf galaxies and the relation to the environment provides us with a key to how different types of dwarf galaxies are emerging. In this review we will put the emphasis on the exotic starburst phase, as it seems less important for present day galaxy evolution but perhaps fundamental in the initial phase of galaxy formation.

\section{Introduction}
Star forming dwarf galaxies (SFDGs) are characterized by low mass, low chemical abundances, high gas and dark matter (DM) content. They constitute one of the most common types of galaxies in the local universe \citep[e.g.][]{2003A&A...408..845D} and increase in importance with redshift. SFDGs reside in low density environments, mostly with lack of massive neighbors \citep{2011arXiv1101.1093W,2011arXiv1101.1301W}.  In this review, with focus on starburst dwarfs, we will consider a range of morphological types - Sm/Im, dI, gas rich low surface brightness galaxies (LSBGs) and blue compact galaxies (BCGs). 

More than 70\% of all galaxies in the local universe are SFDGs \citep{2004AJ....127.2031K}.  Most of these live a quiet family life and consume their gas at a slow pace \citep[e.g.][]{2009ApJ...692.1305L}. This leads to a modest chemical enrichment, in the low mass end probably further diluted by selective metal ejection by stellar winds or infall of fresh gas. Occasionally SFDGs go wild and enter a starburst phase but in most cases eventually return to the quiescent phase without changing their properties radically. Most of the basic characteristics of SFDGs have been summarized in the review paper by \citet{1984ARA&A..22...37G}.

There is no commonly accepted definition of a dwarf galaxy. In the traditional sense \citep{1971ARA&A...9...35H}, a dwarf refers to a galaxy of low luminosity $\sl and$ small size $\sl and$ low surface brightness. Sometimes only one of these criteria is used \citep[e.g.][]{1986PASP...98....5H,1980dwga.work....3T,1997ApJ...481..157P}. Normally a magnitude limit of M$_B\sim -16$ is applied and this is useful under normal conditions. But gas rich galaxies can change the luminosity and surface brightness drastically if they enter into a starburst phase violating both the luminosity and the surface brightness criteria. A mass constraint combined with a relative gas mass fraction is probably the most relevant (but hard to apply) way of characterizing a SFDG. 

As is evident from studies of the local sample of SFDGs \citep[e.g][]{1997RvMA...10...29G,2003ApJ...596..253S,2009ARA&A..47..371T,2011arXiv1101.1093W}, the details of the star formation history (SFH) of SFDGs can be very different from individual to individual. Over long time scales it is controlled by feedback processes \citep{1995AJ....110.2665M} and environmental influences. Starburst galaxies are rare and, in the present epoch, do not significantly influence the evolution of star forming dwarfs as a type \citep{1984ApJ...284..544G,2004MNRAS.351.1151B,2009ApJ...692.1305L}. On the individual level however, starbursts may significantly alter the conditions momentarily although it seems unlikely that a major fraction of the gas can be permanently expelled as a consequence of the supernovae winds and thus transform a gas rich galaxy to a gas poor. An increase in the star formation rate (SFR) with a factor of 2-3 may be sustained over long time periods while it is unclear if an extraordinary increase in SFR can survive over a time long enough to consume a significant fraction of the gas before it is quenched by feedback effects. The major difference between the bursting state and the normal one seems rather to be the mode of star formation, the tendency for stars to form in super star clusters (SSCs) \citep[e.g.][]{1995AJ....110.2665M,2010MNRAS.407..870A} and possibly switch to a flatter IMF \citep{1990ASSL..160..125S,2008ApJ...675.1319H,2011MNRAS.412..979W}.

Why is it interesting to study SFDGs? In many respects these galaxies resemble our notions of the fundamental building blocks of galaxies in the early universe. Very likely, the first dwarf population was the main driver of the cosmic reionization. Searches for Lyman continuum leakage from local SFDGs are therefore important to better constrain the intergalactic ionization field in the early days. 

Simulations of structure formation show that dwarfs merge into more massive units. Locally, we can directly study this process and its effect on the star formation, morphology and gas content. We can learn about star formation processes, shocks and chemical pollution. The lack of shear and the slow rotation of low mass SFDGs open for direct studies of the star formation processes and its evolution over long time scales. The rich variation in star formation histories (SFHs) and the frequent occurrence of SFDGs in the local universe give us a wealth of information in great detail. The low dust content make them transparent and relatively easy to study. Finally, dwarfs help us to understand the nature of dark matter (DM). It is still a puzzle why dwarfs have such a surprisingly low baryon/DM ratio with respect to the cosmic value \citep{2010ApJ...708L..14M}. 

\section{Recent surveys}

Our understanding of SFDGs has drastically improved in recent years thanks to data from space telescopes and large groundbased optical, near-IR and radio telescopes. In particular the data from HST, GALEX and 21-cm surveys have improved our understanding of the formation and evolution of SFDGs and their relation to other galaxy types. The ACS nearby galaxy survey \citep[ANGST;][]{2009ApJS..183...67D} uses data from HST to derive CMDs and detailed SF histories. The survey contains 69 galaxies of which 58\% are dIs, 17\% spirals and the rest dEs. The 11 Mpc \ha and Ultraviolet Galaxy Survey \citep[11HUGS;][]{2004AAS...205.6004L} uses \ha and GALEX images to look at the SF properties of SFDGs in the local volume out to 11 Mpc.  Both these surveys also use Spitzer data. A dedicated Spitzer programme, the Spitzer Local Volume Legacy Syrvey (LVL) has observed 258 nearby galaxies, many of which are supplemented by groundbased images in broadbands and \xha. The Faint Irregular Galaxies (65 dIs) GMRT Survey \citep[FIGGS;][] {2008MNRAS.386.1667B}, is using the Giant Metrewave Radio Telescope (GMRT) to perform volume limited ($<$10Mpc) H I observations of SFDGs.  The H I Nearby Galaxy Survey \citep[THINGS;][]{2008AJ....136.2563W}, at the VLA, includes 34 galaxies, of which about a dozen are SFDGs, in the distance interval 3$<$D$<$15 Mpc. The resolution is 7" and 5 kms$^{-1}$. A similar study, 'Little Things', concerns VLA observations of 41 dIs. Finally, the Arecibo Legacy Fast ALFA (ALFALFA) blind extragalactic H I survey \citep{2005AJ....130.2598G} will survey galaxies at high latitudes out to a radial velocity of 18000 \xkms. It will improve dramatically on the predecessor HI Parkes All-Sky Survey \citep[HIPASS;][]{2001MNRAS.322..486B}. As part of the results the survey will provide us with data of hundreds of dwarf galaxies with H I masses $<$10$^{7.5}$\sma.

\section{Basic structure and morphology}

There is a rich variation in morphologies of SFDGs. The local group galaxy NGC6822 is in many respects a prototype of the normal dIs. It is the nearest SFDG without a massive neighbor. In the central area it contains a rich number of H II regions and star clusters of various ages. On larger scales the stars are embedded in an 1.8 10$^8$\sm H I disk  \citep{2000ApJ...537L..95D}. Young stars trace the H I structure out to large distances \citep{2003ApJ...590L..17K}. Apparently also NGC6822 has been influenced by tidal interaction that has caused tidal tails and enhanced star formation in the outer regions. Older stars are more spherically distributed. This demonstrates that also dIs may have had an initial intense star formation epoch during which the galaxy was taken shape but still in the collapse state. The mean SFR over the past 100 Myr is estimated to 1.4$\cdot$10$^{-2}$\smayr  ~\citep{2011arXiv1101.6051E}. Correcting for He, H$_2$ \citep{2010A&A...512A..68G} and heavy elements, we arrive at a gas consumption timescale of about 19 Gyr, rather typical for its type \citep{2004AJ....128.2170H}.

While the luminosity profiles of dIs follow a simple exponential law, the more enhanced SF activity in the centres of BCGs, possibly linked to dynamical instabilities, show a more complex behavior \cite{1988A&A...204...10K,1995AJ....110.2665M,2001ApJS..133..321C}. BCGs often have morphologies reminiscent of mergers and sometimes experience strong starbursts. In an effort to develop a classification system for BCGs,  \citet{1986sfdg.conf...73L} proposed four different types with the most common one having several star forming regions near the centre, embedded in a regular halo. \citet{1988A&A...204...10K} associated the structure with that of dEs. \citet{1996A&AS..120..207P} found that the luminosity distribution in BCGs in the optical region could be represented by 1) an exponential component at large radii describing an underlying older stellar population; 2) a plateau at intermediate radii and 3) a gaussian component at small radii. Infrared observations reveal in many cases a flattening of the exponential profile \citep{2003A&A...410..481N} towards the centre.  The faint component emerges at very low surface brightness, and supports a high age. The presence of a faint extended red component indicates that the host has similarities to dEs. In some cases, as in  the extremely metal poor SFDG SBS 0335Ð052, we observe faint extended emission that is strongly influenced by nebular emission. After correction, the remaining stellar component agrees with a fairly young population  \citep{1997ApJ...476..698I,1998A&A...338...43P} as if the galaxy formed recently.

\section{Gas content}

A substantial part of the SFDGs have high gas mass fractions.  H I masses range from a few 10$^6$\sm at M$_B\sim$ -10 (represented in the FIGGS sample) up to  a few times 10$^{9}$\sm at M$_B\sim$-20. The relative H I mass fraction varies from about 5\% of the dynamical mass at the high mass end to almost 100\% at the low mass end \citep{2008MNRAS.386.1667B,2001AJ....121.2420S}. The H I disk thickness increases with decreasing luminosity, reaching a mean axis ratio of $<q>\sim$0.6 in the faintest dI galaxies \citep{2010MNRAS.404L..60R}.

\begin{figure}[t]
\centering
\includegraphics[scale=.37]{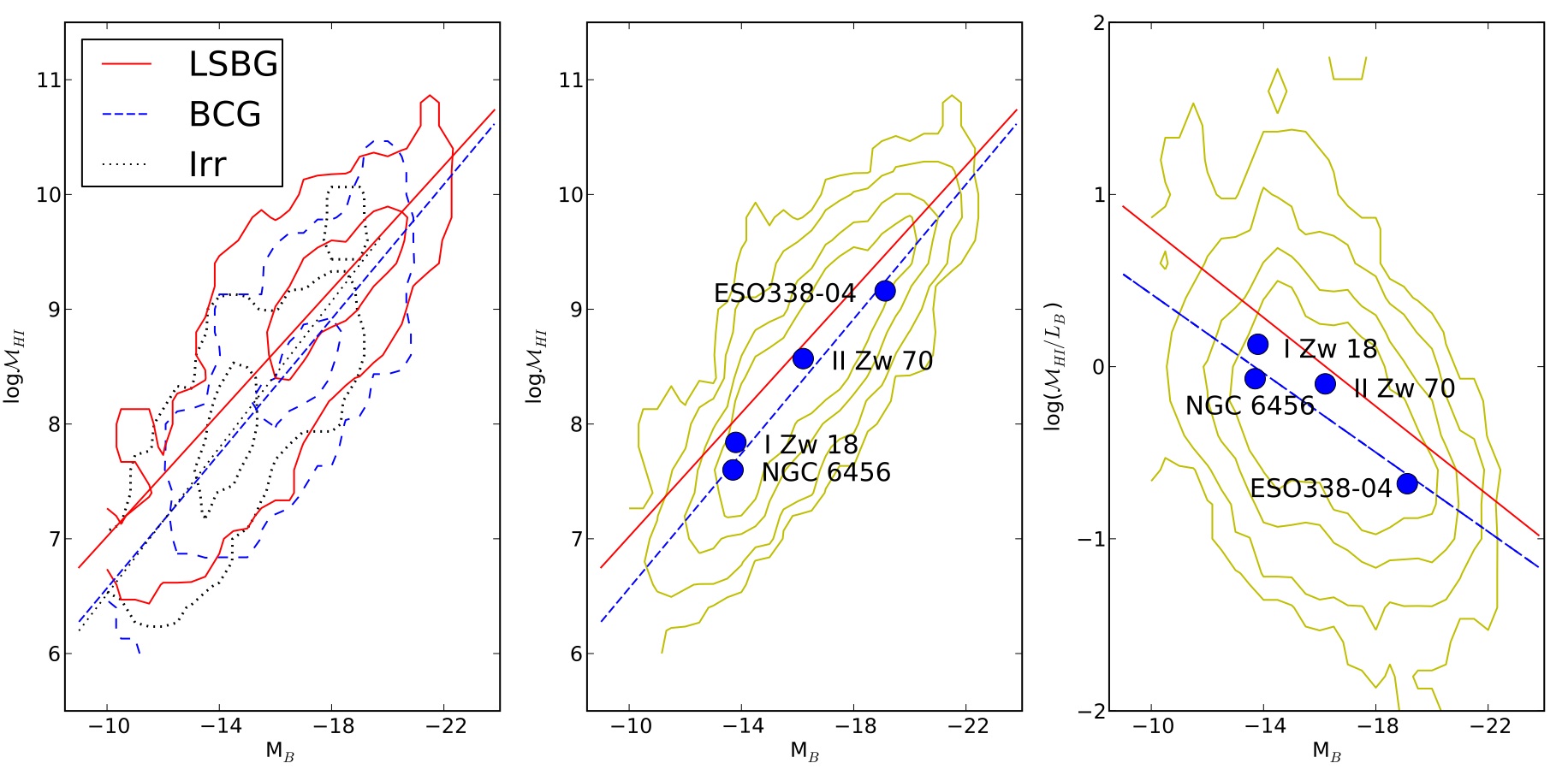}
\caption{{\bf Left:} Isodensity contours of total H I  mass versus absolute B magnitudes for galaxies classified as LSBGs, BCGs and Irrs. The type definition is not strict but varies between different authors (see reference list in running text). Least-square fits to the data sets are shown in solid (LSBGs), dashed (BCGs) and dotted (Irrs). {\bf Centre:} Isodensity contours for the full sample. Also shown are the previous fits to BCGs (blue) and LSBGs (red) and the positions of four well known BCGs. {\bf Right:} log(\mahilbx) vs \mbx. Correlations as before.}
\label{mbhi}       
\end{figure}

In Fig \ref{mbhi} we show the relation between \mahi and \mb based on 199 galaxies loosely classified as Irregular, 315 LSB galaxies and 224 galaxies classified as BCGs or H II galaxies. The sample has been collected from various publications \citep[][Bergvall et al., unpublished]{1997AJ....114.2497V,1998ApJ...496..145L,1999ApJ...511..639I,2001AJ....122.2318B,2002A&A...388..439H, 2002AJ....124..191S,2002A&A...389..405P, 2006ApJ...653..240G,2008MNRAS.386.1667B}. The classification is therefore not homogeneous and the sample is not in any sense complete. But it displays interesting relations. There is a fairly tight linear relation between H~I mass and luminosity for the different types. The \mahilb ~varies with more than one magnitude over the luminosity range.  The spread in luminosity is not significantly different for the different samples. We also see that the distribution of dIs nicely attaches to the faint end of the BCG distribution. It seems that BCGs are merely scaled up versions of dIs. It raises the question about the bursty character of the BCGs. How many galaxies classified as BCGs really differ significantly from 'normal' SFDGs in terms of SFRs and gas consumption rates? One could have expected that the BCGs, if they are bursting, would have an increased dispersion in luminosity since we would catch the galaxy in different stages of a transient event.  But this is not the case. Is it telling s that the BCGs are not bursting but form stars at a sightly higher rate than LSBGs over a long period of time? 

There is an ongoing discussion about the relationships between the morphological types of SFDGs and one important question is how LSBGs and BCGs are related \citep[e.g.][]{2000AJ....120.2975S}. Both have low metallicities and high gas mass fractions and reside in similar environments \citep{2009A&A...504..807R,1995ApJ...443..499P}. If BCGs are bursting, clearly their progenitors must be either irregulars or LSBGs. As the galaxy enters the burst, the luminosity would increase, while the H I mass slowly would decrease. The galaxy would move to the right  in Fig \ref{mbhi}. As we see, the relation for the BCGs is shifted relative to the LSBGs with about one magnitude. Thus, it seems obvious that if LSBGs are the progenitors and the B luminosity roughly can be used as a measure of the SFR, they would increase the SFR with a factor of $\sim$ 2 in the transition. This agrees with results from previous investigations \citep[e.g.][]{1996A&A...314...59P}. However, with such a small change, Occam's razor urges us to primarily consider the case where BCGs always had a higher SFR than the LSBGs. Thus, there is no strong support of starbursts in BCGs from these data. More surprise is found from the positions of three well known BCGs in the diagram, NGC 6456 (see below), I Zw 18, II Zw 70 \citep{2008A&A...477..813K} and ESO 338-04 (=Tol 1924-416, see below). The galaxies all cluster along the mean line. This is unexpected since they are regarded as strong starbursts and therefore should be separated from the distribution of 'normal' BCGs. We would expect strong starburst BCGs to deviate from their candidate progenitors with 2-3 magnitudes. But the shift is only $\sim$ 1 mag. Extinction effects play a minor role since E(B-V)$<$0.05  in all 4 galaxies \citep{1981A&A...103..305L,1978ApJ...226L..11O,1985A&A...146..269B,2010ApJ...721..297M}.

Another clue to the relation between LSBGs and BCGs can be obtained from a study of the spatial distribution of H I. The scalelength versus \mahi  from small sizes up to a few 10 kpc follows that of a constant surface density. But the scalelength also correlates with surface brightness. In a comparison between BCGs, LSBGs and dI galaxies, \citet{2001AJ....122..121V} find that the scalelength of BCGs is systematically smaller than for LSBGs and dIs and roughly corresponds to the optical radius. The central surface density is about a factor of 5 higher. Thus, if LSBGs are progenitors of BCGs, they first must reduce the angular momentum of the gas during the transition phase, in order to increase the H I surface density and thereby the SFR. In a similar discussion, \citet{1992MNRAS.258..334S} concluded that LSBGs and BCGs as classes probably were unrelated.

An important issue concerns the H$_2$ content. Stars form out of molecular clouds and it is known that the relative \hh mass  can be substantial in SFGDs. \hh lacks a dipole moment, making direct observations of \hh extremely difficult. Instead indirect observations of the associated CO molecule is used. The \hh mass is then calculated using the CO-to-\hh conversion factor, \acox. The value of \acox is however very poorly constrained at low metallicities since the CO line becomes weaker and also because the radiation field, especially from starburst regions, tend to dissociate the molecule. The conditions under which this happens cannot be modelled to high accuracy and consequently much of our knowledge about the conversion factor is empirical. The problem becomes evident from observations of galaxies in the local group. \citet{2011arXiv1102.4618L} find that while M31, M33, LMC and the Milky Way have quite similar values, \aco can increase with a factor of 5-10 in SMC and NGC6822. This has serious implications for models of chemical evolution (see below) and timescales of gas consumption, a central issue as we discuss starburst dwarfs. The transition from normal to high values of \aco seems to take place at a metallicity of 12 + log(O/H) $\sim$ 8.2-8.4. Alternative means of estimating the \hh mass have been exploited. Much \hh is residing in photodissociation regions (PDRs). In fact PDRs may be the dominating phase in starburst nuclei \citep{1991ApJ...373..423S}. PDRs contain ions of low ionisation potential, like C$^+$, giving rise to the strong emission line [C II]$\lambda$158$\mu$. The emission is strongly correlated with H$_2$ \citep{1985ApJ...291..755C} and can be used to estimate the \hh mass. As an example we can have a look at IC 10 where \citet{1997ApJ...483..200M} speculate that, as a consequence of the strong UV field and low metallicity, small CO cores are formed, surrounded by extensive [C II] emitting regions containing H$_2$. H$_2$ masses based on the CO flux would therefore be severely underestimated. A similar situation may be at hand in the luminous (M$_B\sim$-21) BCG Haro 11 \citep{2000A&A...359...41B}. No doubt the Herschel observatory will provide us with data to further unveil the interplay between star formation and the ISM.

Although with some scatter, there is a clear one-to-one relation between SFR and molecular mass \citep{2005ApJ...625..763L,2011arXiv1101.1296K}. In more massive systems one may explain this relation as a coupling between the coolant CO (associated with \hhx) and conditions for collapse of giant molecular clouds (GMC). In low mass systems though, a more likely explanation seems to be that the transition from predominantly atomic hydrogen to molecular occurs under the same conditions as a dramatic drop in gas temperature that destabilizes clouds and initiates collapse \citep{2011arXiv1101.1296K}. While the mean ratio between \hh and H I masses increases with metallicity, the scatter is large at low metallicities and there are cases when the \hh mass is more than 10 times higher than the H I mass, like NGC4630 \citep{2005ApJ...625..763L}. Another case is perhaps Haro 11, discussed above. Perhaps there is a yet unknown mechanism that stimulates an overproduction of gas in molecular state, leading to a starburst event.

\section{Chemical abundances}

\begin{figure}
\sidecaption
\includegraphics[scale=.42]{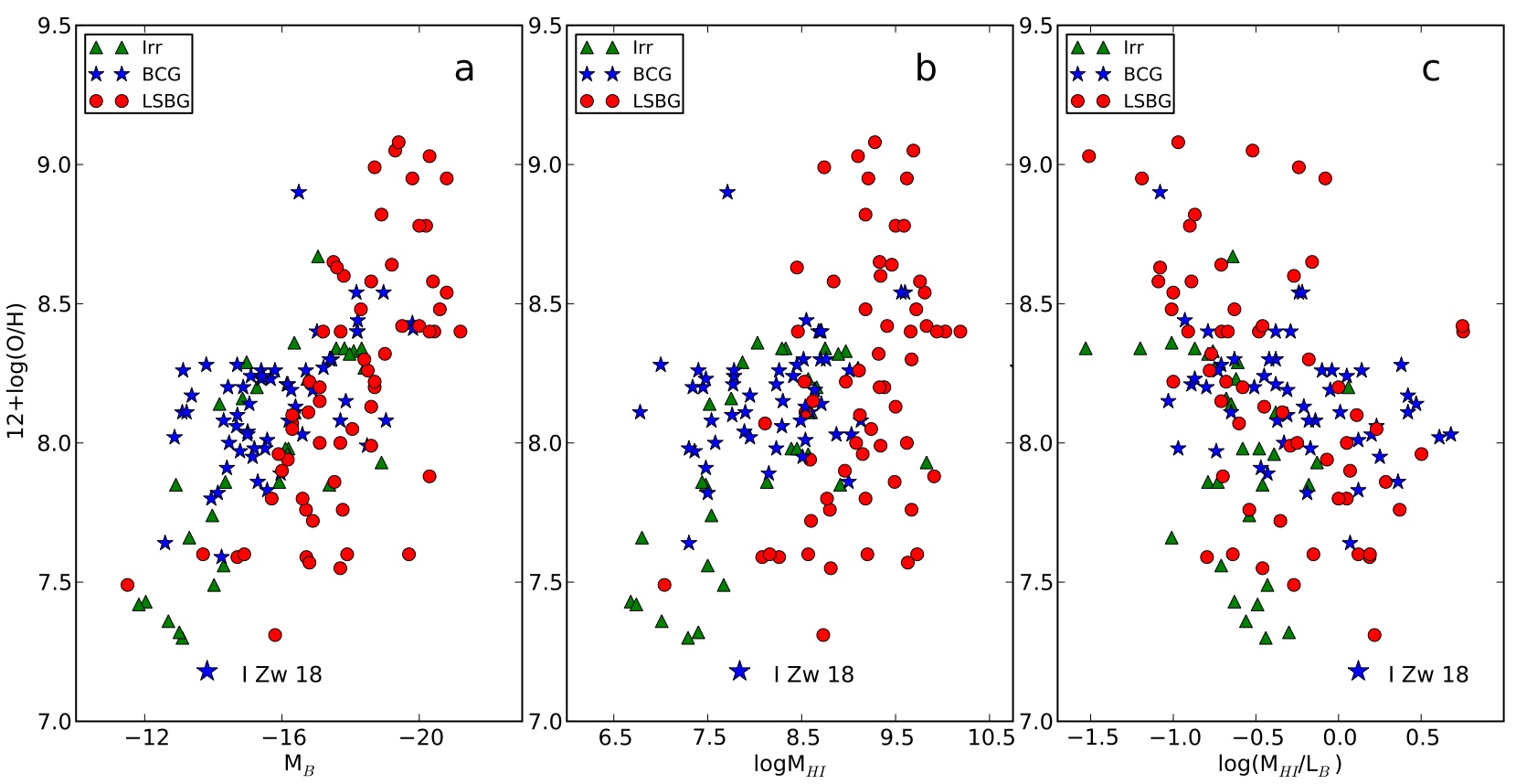}
\caption{Oxygen abundances of Irr, BCG and LSB galaxies as function of absolute B magnitude ({\bf a}), H I mass ({\bf b}) and (H I mass)/(B luminosity) ({\bf c}). The samples in the three diagrams are identical and are obtained from \citep{1994ApJ...420..576M,1995A&A...302..353R,1997AJ....114.2497V,1998ApJ...496..145L,1999ApJ...511..639I,2001AJ....122.2318B,2002A&A...388..439H, 2002A&A...390..891B,2002AJ....124..191S,2002A&A...389..405P,2005A&A...434..887H,2006ApJ...653..240G,2008MNRAS.386.1667B} and additional scattered information in the literature.}
 \label{ohi}occur
\end{figure}

SFDGs are characterized by low chemical abundances. For a review of metal poor SFDGs, see \citet{2000A&ARv..10....1K}. In \hii regions, O/H varies from $\sim$3\% solar to about half solar (using the solar oxygen abundance from \citet{2005ARA&A..43..481A}) for the most massive SFDGs \citep{2002ApJ...581.1019G}. As shown by \citet{1989ApJ...347..875S}, there is a correlation between luminosity and metallicity among local SFDGs. As more data are added though, the scatter increases, in particular if the luminosities are based on emission in the blue part of the spectrum, dominated by young stars \citep[e.g][]{2009A&A...505...63G}. This is partly due to inhomogeneities in the data but cannot be the full explanation as becomes evident if we compare the location of LSBGs in the L-Z diagram with that of BCGs (Fig \ref{ohi}a). Again the data sample is selected in a subjective way and is limited in size since we included only galaxies with both H I masses and O abundances. We see a few striking details. First of all, there is a clear separation between the LSBG sample and the BCG/dI sample. They seem to form two parallel sequences, shifted in luminosity by almost 3 magnitudes. We see this division also in Fig. \ref{ohi}b. A least square fit through the whole sample makes no sense, it is only confusing since apparently we are dealing with two different categories of galaxies. In Fig 2c, showing O/H vs. \mahilb, the two types merge. If the B band mainly reflects the star formation rate in the galaxies, then there is no difference between BCGs and LSBGs as concerns the star formation efficiency per gas mass. The major reason for the difference shown in Fig a and b is probably the scale size and surface density of the neutral hydrogen gas. At the same metallicity the LSBGs are larger, brighter and more extended, making the star formation proceed at a slow speed compared to normal galaxies. Perhaps one could say that the difference in distribution between LSBGs and BCGs+dIs reflects a division line between dwarfs and normal galaxies. Only disk galaxies can produce metal rich galaxies, the others remain below a metallicity ceiling of 12+log(O/H)=8.3. Tidal dwarfs \citep{1999IAUS..186...61D} form an exception to the rule and are found at O abundances $\sim$Ê8.5. Among the brighter galaxies we find a small number of BCGs with higher metallicities, mixed with the LSBGs. Possibly we find progenitors to more luminous (\mb $< \sim$-16) BCGs among the LSBGs but hardly among the fainter BCGs since either the LSBG has to get rid of more than 50\% of the gas or increase the metallicity with a factor of a few. 

Before one draws too strong conclusions based on the L-Z or mass-metallicity diagrams one should consider the influence of mass exchange between galaxy and the environment.  \citet{1994A&A...282..709K} realized the importance of determining the abundances in the neutral H I envelope to better understand the chemical evolution. \citet{2010arXiv1010.2154K} find a correlation between metallicities in the neutral gas and the H II gas, indicating that mixing between the newly produced metals and the H I envelope is taking place. There is however a systematic difference in metallicity between the two states. In the L-Z diagram the most violent starburst dwarfs are found in the lower rightmost wing of the BCG distribution. One of the reasons may be that the stellar population in these galaxies manages to ionize a large part of the gas in the central region and consequently the measured metallicity will be lower than for a milder starburst. The most metal poor galaxies, the low luminosity dwarfs IZw18 and SBS0335-052, have metallicities of a few \% solar.This is close to the metallicity derived for the neutral gas in these and other BCGs using UV spectroscopy \citep{1999ApJ...511..639I,1999ApJ...511..639I,2004A&A...413..131L,2002ApJ...565..941T,2001ApJ...554.1021H,2004ApJ...614..698L}.  It appears as if there is a minimum  plateau defined by the mean metallicity of the intergalactic medium, i.e. log(O/H) $\sim$ -5 \citep{2002ApJ...565..941T,2005ApJ...621..269T}. 

\section{Star formation and stellar content}

Star formation in SFDGs is strongly self regulated. In its simple form it can be quantified as a function of gas temperature and density where the density dependence roughly can be described as a quadratic Schmidt-type law \citep{1959ApJ...129..243S,1969MNRAS.145..405L}. As massive stars are formed, the gas is heated and ionized, striving to quench further star formation. But there is also a positive feedback from star formation that seems to stimulate cloud formation, as demonstrated in the correlation between stellar surface density and the presence of young stars \citep{1998ApJ...493..595H}. Chemodynamical modelling demonstrate that these feedback effects on the order of a few Myr, establishes an equilibrium that stabilizes the star formation \citep{1995A&A...296...99K}. Because of the density dependence, low mass SFDGs consume their gas very slowly, on a time scale larger than a Hubble time \citep{1985ApJS...58..533H,2004AJ....128.2170H}. occasionally, e.g. when galaxies collide and merge, bursts of star formation may occur. All these basic properties of SFDGs were discussed amost 40 years ago by \citet{1973ApJ...179..427S}. Some of the most metal poor galaxies resemble young systems and for a period of time the possibility that these galaxies could be genuinely young was discussed. After a while it became evident that almost all young galaxy candidates hosted a regular component with colours similar to an old stellar population \citep[e.g][]{1985AJ.....90.1457H}. It is now generally believed that, with few exceptions, the minimum age of the oldest population in the SFDGs is at least 1-2 Gyr. Whether this is true or not for the most  widely discussed young galaxy candidate I Zw 18, is still a matter of debate  \citep{2000ApJ...535L..99O,2004ApJ...616..768I,2007IAUS..235...65T} whereas a second young candidate, SBS 0335-052, is still lacking any evidence for the presence of old stars.

While the fundamental principles of local star formation processes can be fairly well understood, the parameters determining the global star formation processes partly have to be determined empirically. We know that small, gas rich SFDGs have consumed less than $\sim$5\% of their present gas mass in star formation \citep{2001AJ....121.2420S}. The relative amount of stars increases with luminosity. We can think of three basic scenarios to explain this. Either the star formation efficiency increases with mass so that most of the stars in massive galaxies were formed at high redshifts while the low mass galaxies have had more or less a constant SFR. This is the closed-box scenario. In the other two scenarios the box is open. We can assume that the star formation efficiencies (SFEs) have been comparable but low mass galaxies have to a larger extent renewed their gas supply through a continuous infall of fresh gas. In the third alternative SFEs are similar but we now assume that the massive galaxies have expelled their gas. Finally we have to be open for the possibility both the SFE and the mass in-and outflows are dependent on both the galaxy mass and metallicity. We can hope to distinguish between the alternatives through observations of the chemical abundances and in observations of the environments around the galaxies.

The SFR in low mass galaxies have been determined using different methods: the UV flux,  the 7.7$\mu$ PAH flux, the FIR dust emission, the radio continuum or the \ha luminosity \citep{1998ARA&A..36..189K,2009ApJ...701.1398S}. While most methods give information about the present SFR,  radio continuum fluxes also give a direct measurement of the SFR during the latest 10$^8$ years \citep{2002AJ....124..675C}.  The most common method however is to use the \ha luminosity, after correction for dust extinction. This works quite well in most SFDGs. At higher masses, and in particular in starbursts \citep{1998ApJ...503..646H}, a substantial fraction of the ionizing radiation is absorbed by dust and extinction corrections are not adequate. Therefore the SFR determined from \ha has to be supplemented by the amount derived from the far infrared. In low mass galaxies with low SFRs or in LSB galaxies, where the star forming regions are sparsely distributed, we notice a significant difference between the SFR determined from \ha as that derived from the UV flux \citep[but see][]{2009ApJ...706..599L,2010arXiv1011.2181L}. These observations and recent analyses of GALEX data \citep{2011ApJS..192....6L} indicate that stochastic effects due to low SFR per area have more impact on \ha than on the UV fluxes, representing a broader range in stellar mass. UV fluxes therefore may be more useful to measure SFR in the low mass - low surface luminosity end.

Using \xha, \citet{1989ApJ...344..685K} studied star formation in disk galaxies and found that the SFRs declined dramatically at a H I surface density of 3-4 \sma/pc$^2$. He derived a value for this critical column density from the Toomre criterion. In a later investigation \citet{1998ARA&A..36..189K} derived a relation between SFR and $\Sigma_{gas}$ over a wide mass range, incorporating dwarfs as well as massive starburst galaxies. He found that the SFR follows the relation nowadays called the Kennicutt-Schmidt law, i.e. SFR $\propto \Sigma_g^{1.4}$, where $\Sigma_g$ is the gas surface mass density.   As a first approximation one may interpret this relation if the SFR is a product between the surface mass density and the collapse time of a supermassive molecular cloud. But Kennicutt also found that the SFR correlated with the rotation period of the disk so there must also be some dynamics involved. How valid are these relations for the SFDGs? In a study of star formation in dI galaxies, \citet{1998ApJ...493..595H} found that $\Sigma_{gas}$ was a factor of $\sim$2 lower than the critical density in disk galaxies. In the FIGGS survery, the low mass end of SFDGs is mapped in H I  \citep{2009MNRAS.397.1435R}. Some of the galaxies investigated, as well as other dIs, have surface densities below the critical limit. The FIGGS survey contains galaxies with surface densities constantly below the Kennicutt threshold. The median H I mass of 2.8 10$^7$\sma and a median blue magnitude M$_B$= -13.2. These galaxies do have lower SFR than the Kennicutt-Schmidt law predicts but star formation is not completely halted. They should be quite relevant as templates for the first generation galaxies in the early universe.

Star formation cannot occur without formation of molecular clouds. All massive molecular clouds collapse and form stars. This is evident from the nearly linear relation between SFR and mass of molecular clouds \citep{2005ApJ...625..763L}. Leroy et al. calculated the ratio between molecular gas mass and K luminosity, B luminosity, FIR luminosity and dynamical mass. Somewhat surprisingly these ratios were not found to vary significantly along the Hubble sequence. The dwarf galaxies in this respect are just scaled down versions of larger galaxies.  The molecular mass is proportional to the stellar mass. The ${\sl M} _{mol}$/${\sl M} _{mol+H I}$ ratio however, increases with mass. Thus, the more massive galaxies are more efficient in forming molecular clouds, probably as an effect of increasing pressure with gas mass. The resulting increased density in the gas governs the formation of molecules. The scatter in the  SFR-${\sl M} _{mol}$ relation however increases with decreasing mass and as mentioned above the H$_2$ content of low mass SFDGs is still an open issue.

In previous sections we discussed the star formation activity in different SFDG types based on the B luminosity. A more accurate information be be obtained from \xha. \citet{2004AJ....128.2170H} investigated over 100 SFDGs and compared the SFR between galaxies of types classified as Im, BCDs, Sm and Sab-Sd galaxies. A prominent difference between BCDs and the other types is the high concentration of star formation towards the centre in BCGs and that there is a much stronger gradient in the ratio between the \ha and V luminosities.  This indicates that gas recently migrated from the outer regions to the inner within one gigayear. As a consequence, the gas surface densities have increased. Consequently, applying the Kennicutt-Schmidt law, the star formation rate per gas mass has increased and the surface brightness is higher. The star formation efficiency, however, does not seem to be significantly higher in BCDs compared to the other types. 

In the 11HUGS survey, $\sim$300 SFDGs within the 11 Mpc distance are imaged in \ha and UV. An important result is that most of the stars formed in these galaxies are formed in a quiescent mode. Only about 1/4 are formed in bursts and only a few \% of the galaxies are now in a bursting mode (in their study a burst was defined as a galaxy having an equivalent width in \ha, EW(\xha) $>$ 100\AA). This result is in agreement with the star formation properties of galaxies in the SDSS where it is argued that 20\% of the star formation occurs in bursts \citep{2004MNRAS.351.1151B}. 

This modest, well regulated SF activity is confirmed also in more detailed studies of galaxies the backyard universe. Recently a study of colour-magnitude diagrams of 60 local dwarf galaxies from the ANGST project was presented \citep{2011arXiv1101.1093W}. Almost 80\% of these are SFDGs. The galaxies range in M$_B$ from -8.23 to -17.77 and in distance from 1.3 to 4.6 Mpc. It was found that on average, the typical dwarf galaxy formed 50\% of its stellar mass by z $\sim$ 2 and 60\% by z $\sim$ 1. Much of the differences between the different morphological types (including dSph) occur during the latest 1 Gyr and is strongly environmentally dependent. Stellar feedback seems to have less importance for the transition between the different types. Instead, a mechanism of the type 'tidal stirring'  \citep{2001ApJ...547L.123M,2001ApJ...559..754M,2006MNRAS.369.1021M} seems to be more attractive. A simple description of the SF history, as e.g an exponentially decaying SF, appears to be inconsistent with the observations. A more complex SF history is needed.

\section{Starbursts}
The name Blue Compact Galaxy was coined by Fritz Zwicky \citep{1965ApJ...142.1293Z} due to their blue colours and compact appearance on photographic plates. About 10 years later we got access to observational capacities good enough to make it possible to study these galaxies in detail.  It was found that indeed the galaxies had high gas masses and low metallicities, as expected if they were young. In 1972-73 Searle, Sargent and Bagnuolo applied a model of the spectral evolution of galaxies on observations of late type galaxies \citep{1972ApJ...173...25S,1973ApJ...179..427S}. Among these were a few of the bluest. Searle et al. addressed the question of youth and concluded that probably the galaxies were not young but experienced short bursts of star formation.  Prior to this \citet{1978ApJ...219...46L} came to the conclusion that starbursts were generated by interaction and mergers. According to their study, starburst should be a common phenomenon and a characteristic consequence of close encounters and mergers. This generated a new mode of interpretation of star forming galaxies and in 1980 \citep{1980ApJ...238...24R} the word starburst was accepted. A few years later starburst galaxies had become one of the hottest topics in extragalactic astronomy. During the 2000s, about 800 papers per year contain the word starburst in the abstract. 

At the same time as this outstanding burst of interest in starbursts occurred, there was also a growing confusion concerning starbursts. What was it? How should a starburst be defined? This is a persistent problem \citep{2009ApJ...698.1437K}. Classically, it was considered that the star formation rate in a starburst should be significantly enhanced compared to the normal SFR for that specific morphological type. Alternatively, the gas consumption time scale should be significantly smaller than a Hubble time. Gradually there was a need to subdivide the starburst population into global starbursts and nuclear starbursts. The latter occur almost exclusively in normal (non-dwarf) galaxies in a local $\sim$ kpc sized region in the centre. Many workers use  EW(\xha) as a measure of a starburst. One rather widely used criterion for a starburst is EW(\xha)$\geq$100 \AA. According to models \citep{2001A&A...375..814Z} however, a galaxy with constant SFR will have EW(\xha)$\approx$100\AA~ after a Hubble time. Thus the EW(\xha) criterion is not optional. A better starburst indicator is probably the birthrate parameter, $b$ \citep{1986FCPh...11....1S,1998ARA&A..36..189K}. It is defined as the ratio between the present SFR and the mean past SFR: $b = \frac{SFR}{<SFR>}$. A galaxy with $b>$3 can be regarded as a mild starburst galaxy while one with $b>10$ is of the 'classical' starburst type. In the study of star forming galaxies in the SDSS, \citet{2004MNRAS.351.1151B} concluded that about 20\% of star formation takes place in galaxies with b=2-3. It is a matter of taste if one wants to call this a starburst but the starburst concept gets very watered-down if all galaxies with b$>$2 are to be regarded as starbursts. If we use the strong starburst criterion, $b>$10, only 3\% of the star formation takes place in bursts.

\begin{figure}
\sidecaption
\includegraphics[scale=.6]{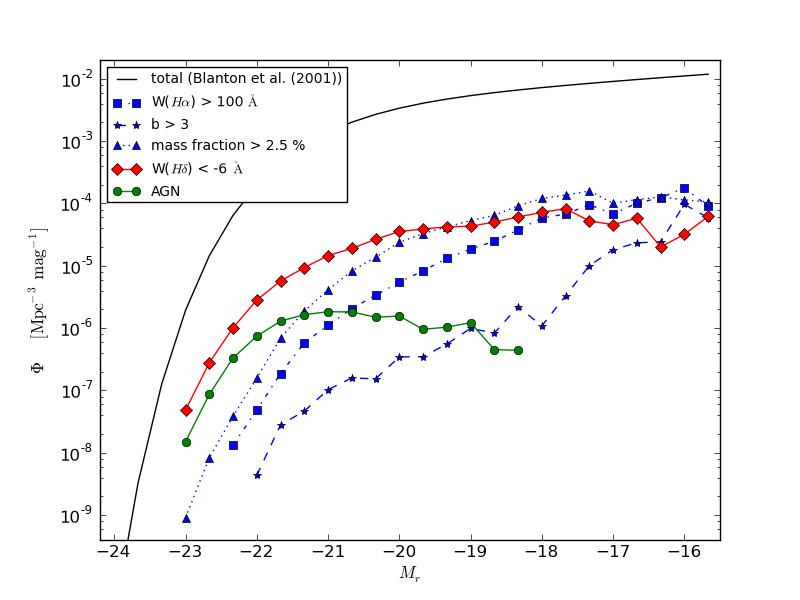}
\caption{The galaxy luminosity function as based on the SDSS (solid line, from \citet{2001AJ....121.2358B}. Also shown are the luminosity functions of galaxies with EW(\xha) $>$100\AA ~(squares), $b>$3 (*), mass fraction$>$2.5\% (triangles), post-starburst galaxies (diamonds) and AGNs (dots).}
 \label{lumf}
\end{figure}

We (Bergvall et al. 2011, in proc. of  'Star forming dwarf galaxies', Crete 2008) carried out a similar study as Brinchmann et al. but with different models and a supplementary set of data of post-starburst galaxies. We modelled the stellar population simply as a mixture of an old stellar population and a young burst. In Fig \ref{lumf}, we show the luminosity function for different subsamples of SFDGs in SDSS, DR7.  As concerns the relative number of galaxies with EW(\ha)$>$100\AA, our results agree with the results from the 11HUGS survey, i.e. that these constitute a few percent of galaxies in the luminosity range -17$\leq$M$_B<$-15. If we apply $b>3$ as a condition for starbursts, we find that it is fulfilled by $<$1\%  of the sample. We also find that the $b>3$ criterion is almost fulfilled for galaxies with EW(\xha)$>$100\AA ~at the low luminosity end of the distribution. As we go towards higher luminosities the EW(\xha) criterion is loosing relevance as a starburst criterion.

\begin{figure}
\sidecaption
\includegraphics[scale=.45]{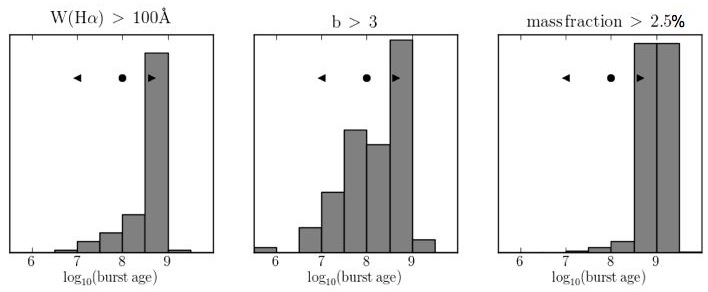}
\caption{Results from modelling of SDSS spectra of SFDGs. The model is a simple mixture of an old and young stellar population. The histograms show the resulting ages of the young component under different conditions. The marks indicate starburst durations derived from other groups using other data. Among these are the results from \citet{2010ApJ...724...49M} (right pointer), \citet{2009ApJ...692.1305L} (dot) and various results arriving at short durations discussed in \citet{2009ApJ...695..561M} (left pointer).}
 \label{age}
\end{figure}

It is interesting to compare these results with studies of nearby galaxies. In the low mass end, the starburst concept more or less looses its meaning. Stochastic effects dominate the star formation history and single star forming events can change the galaxy luminosity significantly. Since the typical dwarf galaxy formed 40\% of its stellar mass at z$<$2, a birthrate parameter of $b$=3 means that the SFR increases with a factor of 6 with regard to the mean SFR from z=2 (3.3 Gyr).  Therefore it may be well motivated to redefine the criterion. Thus, in their study of nearby SFGDs, \citet{2009ApJ...695..561M,2010ApJ...724...49M} accordingly consider only the latest 6 Gyr when they discuss starburst properties, $b$=$b/<b>_{(0-6Gyr)}$. They also prefer to lower the condition further and consider the galaxy to be in a starburst condition if this modified $b$ is $>$ 2. Among their targets, NGC 6456 has the record, $b$=7.6. We added this galaxy to the Fig. \ref{mbhi}. Surprisingly, the galaxy deviates very little from the mean relation for BCGs. Quinn et al. derive the star formation history from CMDs. The find that the duration of the starburst in these 18 galaxies is surprisingly long, typically 450-650 Myr. This is three times the duration adopted by \citet{2009ApJ...692.1305L} and indicates that self-quenching of the starburst is a rare phenomenon. McQuinn et al. continues to discuss other similar studies of the duration of starbursts and cite several independent investigations that arrive at much shorter durations, often $<$ 10 Myr. In Fig \ref{age} we have a look at the durations we derived for the SDSS sample.  If we apply the EW(\xha)$>$100\AA~criterion, the duration of this period agrees with what McQuinn et al. found. If we apply the birthrate criterion, the duration we obtain is inbetween the McQuinn value and the Lee et al. value. What is also interesting to see in the diagram is that starbursts consume a minute fraction of the available gas. Mass consumption rates higher than a few \% are only found among the very longlived "bursts".

From an historical point of view is becomes clear that the starburst concept now has become rather watered-down. The problem becomes apparent when we go to the most extreme starburst dwarfs where we may have $b_{(0-14Gyr)}>$15-20, i.e. with a specific SFR about 20 times higher than starbursts classified with the weak criterion. SFDGs with $b_{(0-14Gyr)}>$3 are rare, constituting $<$ 1\% of all dwarf galaxies. Although these galaxies are involved in a starburst at the moment, it does not necessarily mean that the starburst will influence the evolution of the galaxy significantly. Neither the mass consumption during the burst or the effects of the stellar winds may be important. Let us look at two examples, ESO 338-04 and Haro 11. ESO 338-04 is interacting with a neighboring galaxy \citep{2004ApJ...608..768C}. It has a $b$ parameter of 11 \citep{2001A&A...374..800O} and produces myriads of clusters/super star clusters \citep{1998A&A...335...85O}. The formation of massive star clusters is a common phenomenon in starburst galaxies and is explained as an effect of high pressure \citep{1997ApJ...480..235E}. What is surprising is that in Fig \ref{mbhi}, the galaxy looks completely normal. Since the $b$ parameter is a comparison between the past star formation and the present, the past star formation may have been very modest. Then we can obtain a high $b$ parameter even if the SFE is normal in relation to the available gas mass. 

The other example is Haro 11, mentioned above in connection to the \hh discussions. It has no neighbors but morphologically and kinematically the galaxy appears to be involved in a merger \citep{1999A&AS..137..419O,2001A&A...374..800O,2008ApJ...677...37O}. It has a gas consumption timescale of $\sim$ 100 Myr a birthrate parameter $b\sim$ 15 and is involved in a vigorous global starburst, producing a large number of super star clusters \citep{2010MNRAS.407..870A}. The cluster formation efficiency is about 10 times higher than in large spiral galaxies. The FIR SED of Haro11 is abnormal \citep{2008ApJ...678..804E}. This is probably due to a very young stellar population and suggests that stellar winds should be important. A strong superwind \citep{2006A&A...448..513B,2007ApJ...668..891G} is observed from the central region and Chandra observations \citep{2007ApJ...668..891G} confirm that this wind is penetrating the gaseous halo and may open channels to the halo. In this context it is interesting to compare Haro 11 to high redshift galaxies, in particular Lyman break galaxies \citep{2007ApJ...668..891G}.  There have been claims for and against a Lyman continuum leakage in the galaxy \citep[][Leitet et al., submitted]{2007ApJ...668..891G} and therefore Haro 11 may also be a proxy for Lyman continuum galaxies in the era of cosmic reionization. Today there is a strong support for the scenario that SFDGs  were the main drivers of the cosmic reionization \citep[e.g.][]{2007ApJ...670..928B,2008ApJ...673L...1G,2010ApJ...710.1239R,2011Natur.469..504B} and there are indications \citep{2009ApJ...706.1136O} that the dwarf galaxies at high redshifts are both metal poor, have a flatter IMF and a higher escape rate than local galaxies. But from studies of local metal poor dwarfs that are leaking, we can better understand the physical conditions necessary for leakage.

\section{Additional evolutionary aspects}

Strong starbursts typically have low ages. Therefore we expect to be able to trace the history of this dramatic event. What triggered the burst? \citet{2004AJ....128.2170H} found no support for tidally triggered starbursts in their sample of BCGs. Several strong starburst dwarfs live in isolation but show signs of multiple nuclei, unordered velocity fields, morphological large-scale distortions etc which supports mergers as the most common triggering mechanism \citep[e.g.][]{2001A&A...374..800O}. Understanding the triggering mechanism in low mass starbursts is more of a challenge since the turbulent velocities caused by stellar winds are comparable to the gravitational effects and the possible merging components have no prominent central concentrations.

A burning question is how the different types of dwarf galaxies are related and the possible transition from one type to another. This is an ongoing discussion since the 1970s \citep{1973ApJ...179..427S,1983ApJ...268..667T,1986ApJ...309...59L,1988MNRAS.233..553D,1989ApJ...339..812H,1991AJ....101...94D,1994MNRAS.269..176J,1996A&A...314...59P,1996A&AS..120..207P,2001AJ....122..121V}. It is clear of course that gas rich galaxies in isolation turn into gas poor. But the reverse should not be excluded. There are various possibilities of how this could be achieved except for pure gas consumption in star formation. In starbursts, gas may be expelled as consequence of winds from massive stars and supernovae. Ram pressure stripping as the galaxy falls in towards a cluster with a higher density of the intergalactic medium is another option. Harassment and tidal stirring have also been proposed as important mechanisms. The first step in these studies is to make a comparative investigation of the stellar populations and their distributions in gas-rich vs. gas-poor galaxies.

\citet{1996A&A...314...59P} compare the structural properties of different dwarf galaxy types. The BCDs typically has a central surface brightness 1.5 mag. brighter and a scale length a factor of 2 smaller than dEs and dIs. They conclude that there may be evolutionary connections between BCDs, dEs provided the BCD modifies its structural properties during the transition by a change in the potential field as a result of e.g. a global in- or outflows of gas \citep{1974MNRAS.169..229L}. Outflows of gas in galactic superwinds \citep{1993ASSL..188..455H} is found in many starburst galaxies \citep[e.g.][]{1990A&A...236..323I,1994A&A...291L..13P,1995ApJ...438..563M,1995A&A...301...18L,1996ApJ...458..524I,1996AJ....112..146Y,1998A&A...334...11K,1999A&A...349..801M,2001MNRAS.327..385S,2001ApJ...554.1021H} and the energetics cope with those needed to expel the gas. In some SFDGs it has been argued that outflows have been efficient mechanisms to remove the gas \citep[e.g.][]{2000MNRAS.317..697E,2007ApJ...668..891G}. This has been supported by some models of SN driven gas outflows \citep{1986ApJ...303...39D,1999ApJ...513..142M,2000MNRAS.317..697E,2001ApJ...554.1021H}. From a cosmological point of view, dwarf galaxies are remarkably devoid of baryons \citep{2010ApJ...708L..14M} and it has been argued that outflows could explain this condition \citep{2003MNRAS.343..249S}. E.g. \citet{2002ApJ...581.1019G}, \citet{2004A&A...425..849P}, \citet{2004ApJ...613..898T} and \citet{2007ApJ...658..941D}  all find that the effective yields are lower than expected in a closed-box scenario, supporting an ejection of metals. Dalcanton argues however that the amount of mass lost need not be more than 15\%. In the dwarf starburst Mrk 33 \citet{2001MNRAS.327..385S} find support for ejection of most of the produced  metals and a small amount, ~1\%, of the ISM. These findings are in conflict with other results  \citep{2001MNRAS.323..555L,2006ApJ...647..970L} which find support for the closed-box model. A potential problem when investigating the validity of the closed box model is to derive reliable gas masses at low metallicities, in particular molecular hydrogen. The galaxies producing the most metals are also supposed to have the largest amounts of molecular hydrogen and therefore the total gas mass, and effective yield, may be underestimated. 

While the question about selective metal outflows is still unsettled there is little support of SN blowouts of large fractions of the ISM both from observations and recent modelling \citep{2006A&A...445L..39R}. No large sinks of gas, either neutral, molecular or ionized are found in the intergalactic medium in groups  \citep[e.g][]{2001A&A...373..485D,2003AJ....126.2774H} and when outflows are observed, it appears difficult to transfer the gas to the IGM environment \citep[e.g.][]{1999MNRAS.306...43S,2003MNRAS.342..690S}.  Rather, to explain the remarkably low baryon content in dwarf galaxies, X-ray observations \citep{2010ApJ...719..119D} seem to favour a baryon loss due to preheating. A second possibility is that large quantities of baryons never was involved in galaxy formation but reside as warm-hot gas in the large scale cosmic walls \citep{2010ApJ...714.1715F}. The scatter in the fractional gas/star content however is increasing with decreasing mass so at the lowest masses, SN winds or other mechanisms may play a fundamental role.

So, what are then the dominating mechanisms that  can explain a relationship between more massive dwarf galaxies? What will happen to a BCG after a bursting phase? \citet{1998MNRAS.300..705M} find that the DM halos in two BCGs is $\sim$ 10 times higher than in dIs. A high central mass density is found also in Haro 11 \citep{2009arXiv0901.2869C} which makes it unlikely that BCGs can transform into dIs. It seems more probable that BCGs could develop into dEs \citep[e.g.][]{1996A&A...314...59P,2004A&A...419L..43O,2005ApJS..156..345G,2007A&A...474L...9M}. An attractive mechanism to reduce the gas content seems to be the scenario of tidal stirring, tested by \citep{2001ApJ...547L.123M} on the local group galaxies. But, while it appears attractive in an group environment it will not work on isolated BCGs so the problem remains.

\section{Final remarks}
With the advent of new instrumentation we expect a strong progress in the research of star forming dwarf galaxies. We already mentioned the recent/ongoing surveys with HST, Spitzer, GALEX and radio observatories. The Herschel observatory will provide information about heating and cooling processes of gas and dust in low metallicity environments. This will also help us to better understand the important question of  \hh in metal poor SFDGs. LOFAR will eventually detect low-frequency radiation from halos of starburst dwarfs. Using wide field 2-D spectroscopy we will study the nature and triggering mechanisms of starbursts. The conditions for Ly$\alpha$ emission and the coupling to Lyman continuum leakage will be explored and applied on starburst dwarfs at high redshifts \citep{2011ApJ...730....8H}. Gradually we will understand the importance of SFDGs in the cosmic reionization and formation of massive galaxies, the regulating mechanisms of starbursts, the violent star formation processes, the mass exchange with the environment  and the relation between different types of dwarf galaxies.

\begin{acknowledgement}
I thank the organizers of this workshop who invited me to give this talk. In particular I would like to thank Polychronis Papaderos for creating such a nice atmosphere, stimulating social contacts and scientific discussions. I would also like to  thank him and Thomas Marquart for many enlightening and stimulating discussions about the subject of star forming dwarf galaxies. Thomas is also acknowledged for helping me with some of the graphical ingredients of this paper.
\end{acknowledgement}

\bibliographystyle{spmpscinat}

\bibliography{biblio}

\end{document}